# Innovative Integration of 4D Cardiovascular Reconstruction and Hologram: A New Visualization Tool for Coronary Artery Bypass Grafting Planning

**Running title: Advancing CABG Planning with 4D Holography**

Shuo Wang[1#] Ph.D. Candidate, Tong Ren[2.3#] MM, Nan Cheng[2] MD, Li Zhang[1] PI, Rong Wang[2*] MD.

[1] Department of Engineering Physics, Key Laboratory of Particle and Radiation Imaging, Ministry of Education, Tsinghua University, Beijing 100084, China.

[2] Department of Adult Cardiac Surgery, Senior Department of Cardiology, The Six medical center of PLA General Hospital. Beijing 100048, China.

[3] Chinese PLA Medical School, Beijing 100089, China.

[#] Shuo Wang and Tong Ren contributed equally to this work, both as the first authors.

[*] **Corresponding author:**

Rong Wang,

Department of Adult Cardiac Surgery, Senior Department of Cardiology,

The Six medical center of PLA General Hospital,

No.6 Fucheng Road, Haidian District, Beijing 100048, China.

Email: rong_wwang@outlook.com

Tel: +86-15801371359.

# Abstract

**Background:** Planning for coronary artery bypass grafting (CABG) necessitates advanced spatial visualization skills and consideration of multiple factors, including the depth of coronary arteries within the subepicardium, calcification levels, and pericardial adhesions.

**Objective:** This study aimed to address these requirements by reconstructing a dynamic cardiovascular model, displaying it as a naked-eye hologram, and evaluating the clinical utility of this innovative visualization tool for preoperative CABG planning.

**Methods:** We used preoperative 4-dimensional cardiac computed tomography angiography (4D-CCTA) data from 14 patients scheduled for CABG to develop a semi-automated workflow. This workflow enabled time-resolved segmentation of the heart chambers, epicardial adipose tissue (EAT), and coronary arteries, complete with calcium scoring. Methods for segmenting cardiac structures, quantifying coronary calcification, visualizing coronary depth within EAT, and assessing pericardial adhesions via motion analysis were incorporated. These dynamic reconstructions captured spatial relationships, coronary stenosis, calcification, and depth in EAT, as well as pericardial adhesions. Dynamic cardiovascular holograms were then generated and displayed using the Looking Glass® platform. Thirteen cardiac surgeons assessed the utility of the holographic visualization tool on a Likert scale. Additionally, a surgeon visually scored pericardial adhesions using the holograms of all 21 patients (including seven undergoing secondary cardiac surgeries) and compared these scores with actual intraoperative findings.

**Results:** Cardiac surgeons highly rated the visualization tool for its utility in preoperative planning, with a mean Likert score of 4.57/5.0. The hologram-based scoring of pericardial adhesions showed a strong correlation with intraoperative findings (correlation coefficient: $r = 0.786$, $P < 0.001$).

**Conclusion:** This study delineates the structural framework of a visualization tool specifically designed for preoperative CABG planning. It produces high-quality, clinically relevant, dynamic holograms from patient-specific volumetric data, with clinical feedback confirming its practicality and effectiveness for preoperative surgical planning.

# Introduction

Coronary artery disease (CAD) is one of the leading causes of death worldwide[1]. Coronary artery bypass grafting (CABG) is the most frequently performed cardiac surgery globally[2]. Compared to percutaneous interventions or medication therapy alone, CABG provides a survival advantage for patients with complex coronary artery disease[3]. Accurate identification of the coronary arteries and careful selection of the anastomotic site are critical for surgical success. If the chosen anastomosis site is located too deep within the thoracic cavity, the limited operating space may hinder surgical maneuverability. Furthermore, when a coronary artery lies deep within epicardial adipose tissue (EAT) or within the myocardium, intraoperative localization becomes challenging. This can lead to longer operative times, increased tissue trauma, and a reduction in the morphological quality of the anastomosis. In addition, plaque may obstruct suturing, and pericardial adhesions can increase the risk of injury to the heart or major vessels, further prolonging the procedure and raising its overall risk. Therefore, collecting detailed information on pericardial adhesions, spatial orientation, coronary artery depth within EAT, stenosis, and calcification is crucial for preoperative CABG planning.

Enhanced 3D cardiac models that incorporate multiple functional imaging modalities provide more comprehensive insights[4]. A practical strategy to realize precision medicine in cardiovascular care is through the digital twinning of the cardiovascular system, utilizing coronary angiography, echocardiography, and coronary CT angiography (CTA) data[5]. Compared with ultrasound and magnetic resonance imaging (MRI), CTA offers superior spatial resolution and is widely used for reconstructing patient-specific 3D models. While MRI techniques such as 4D cardiac MRI and cine multislice MRI offer excellent temporal resolution along with good spatial resolution and are well-established in clinical practice, 4D cardiac CT imaging offers sufficient spatial and temporal resolution to support advanced visualization. This enables dynamic and functional assessments of cardiovascular conditions. Although 3D and 4D visualization technologies with effective depth cues are already established in clinical practice, enabled by robust 3D viewers that allow simultaneous multi-view display and free-angle reformatting of volumetric datasets, our study hypothesizes that holographic displays may provide additional benefits for specific surgical planning tasks, warranting further investigation.

Building upon the established foundation of clinical 3D visualization, our research explores whether holographic displays can provide complementary benefits for CABG planning. Rather than suggesting limitations in current clinical displays, we aim to explore whether the stereoscopic nature of holography offers added value for specific surgical planning scenarios. Effective communication among healthcare team members can be hindered by individual differences in spatial visualization abilities. Researchers have made significant progress in representing medical 3D images through virtual and augmented reality technologies, as well as various stereoscopic display systems. In recent years, holographic display technology has made notable advances in the fields of surgical planning and intraoperative navigation, with the Microsoft HoloLens emerging as the leading AR/MR headset[6–10].

Holography is widely considered the most advanced form of display technology, as it incorporates all human visual cues, including stereopsis and accommodation, allowing for naked-eye visualization[9]. Although modern head-mounted devices have become more lightweight and support networked multi-user viewing, they still present significant limitations. One persistent issue with virtual reality (VR) is "cybersickness"—a condition marked by symptoms such as headaches, nausea, and vomiting[11]. Additionally, these devices often fall short in providing true depth perception[10]. The HOLOSCOPE™ is the world's first medical-grade holographic system. Elchanan Bruckheimer et al. demonstrated the creation of real-time, interactive 3D digital holograms without the need for headgear, using actual patient voxel data from 3D rotational angiography and intraoperative real-time 3D transesophageal echocardiography in patients with structural heart disease for preoperative assessment[10,12]. Looking Glass Factory has introduced the world's first group-viewable holographic displays[13], offering a significantly more affordable alternative to the RealView HOLOSCOPE™, which typically costs hundreds of thousands of dollars. In contrast, entry-level Looking Glass devices are priced at just a few thousand dollars. These displays offer substantial advantages for simultaneous multi-user viewing, collaborative discussion, and cost-effectiveness, making them particularly valuable for surgical planning in procedures like CABG, which often require input from multidisciplinary teams.

Building on a strong medical-industrial collaboration, our study addresses critical challenges in preoperative CABG planning by designing a 4D cardiovascular hologram based on 4D-CCTA data and evaluating the feasibility of using this holographic visualization tool for surgical planning.

# Materials and Methods

## Study design and overview

A total of 21 patients scheduled for cardiac surgery were evaluated using 4D-CCTA prior to the procedure. This cohort included 14 patients with complex coronary artery lesions who were candidates for CABG, with a mean age of 53.6 years; 13 of them were male. Because pericardial adhesions are common in patients undergoing repeat cardiac surgery, we selected seven patients undergoing secondary procedures as a control group for evaluating pericardial adhesions. These patients had a mean age of 55.5 years, and six were male. Operative reports confirmed extensive pericardial adhesions in all patients undergoing reoperation, and the absence of adhesions in all patients undergoing first-time surgery. All reoperations involved heart valve procedures, with the interval between the initial and repeat surgeries ranging from 1 to 35 years. Among these patients, three had valve dysfunction following corrective surgery for congenital heart defects; one had infective endocarditis after left ventricular outflow tract unblocking, mitral valve replacement, and CABG; one developed infective endocarditis following mechanical valve replacement; one experienced tricuspid insufficiency after left heart valve surgery; one had mitral valve insufficiency after mitral valve plasty; and one had mitral valve insufficiency following prior left heart valve surgery. Inclusion criteria for the study were: (1) patients with complex coronary lesions meeting the clinical indications for CABG and no history of previous cardiac surgery; (2) patients undergoing secondary surgery in whom the autologous pericardium had been closed during the initial procedure. Exclusion criteria included poor-quality imaging data.

Our study aimed to construct a 4D cardiovascular hologram using individualized patient imaging to address key challenges in the preoperative planning of CABG. Clinicians will evaluate the effectiveness of the holographic visualization tool by comparing it with patients' preoperative and intraoperative data, in order to assess the system's practical utility for surgical planning. The surgeon will not use the holograms to guide the procedure or make clinical decisions based on these images.

## Hologram visualization system

The holographic display used in this study is the 8.9-inch Dev Kit from Looking Glass Factory Inc. (Brooklyn, New York, USA), which leverages advanced light field technology to produce highly realistic 3D visualizations of cardiovascular structures. This system generates true volumetric imagery by projecting 48 distinct horizontal perspectives of the same visual content through calibrated lenticular arrays, allowing multiple observers to simultaneously view and interact with the 3D cardiac models from various angles without positional restrictions (Fig 1).

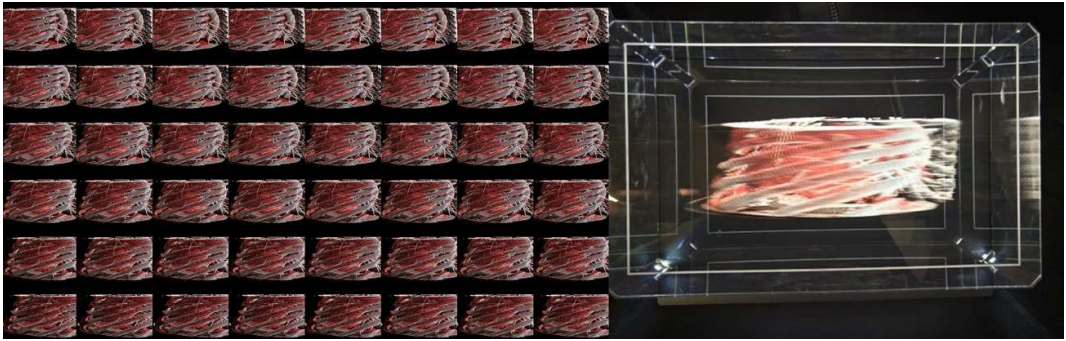

**Figure 1** Holographic display system consisting of 48 viewing angles.

The dynamic cardiac hologram is displayed at 60 frames per second, closely replicating the visual fluidity of normal cardiac cycles. The system provides a horizontal viewing angle of 58°, enabling up to five observers to simultaneously view the same holographic image from different perspectives without distortion. Depth perception is significantly enhanced by the system's ability to render between 45 and 100 discrete depth planes along the z-axis, producing a seamless sense of depth that is crucial for interpreting complex cardiac anatomy. These complementary features—multiple horizontal perspectives and discrete depth planes—combine to deliver a truly immersive 3D visualization experience.

Unlike conventional virtual reality or augmented reality solutions, this technology does not require head-mounted displays or specialized eyewear, allowing surgical teams to communicate naturally while examining patient-specific anatomy. Interaction with the holographic model is enabled via a Leap Motion controller, which supports intuitive, gesture-based manipulation. Users can rotate the model along any axis, zoom into regions of interest, adjust playback speed of the cardiac cycle, and selectively display or hide specific tissue layers—including epicardial fat, coronary vessels, and calcification sites—using natural hand movements. This gesture control system is particularly valuable during preoperative planning sessions, where multiple specialists can collaboratively examine and manipulate complex anatomical structures while maintaining full situational awareness and unimpeded lines of communication.

## Images and processing

CT images were acquired during end-expiratory breath-hold and in sinus rhythm using a GE APEX CT scanner (Revolution Apex, GE Healthcare, Milwaukee, MI) with a retrospective ECG-gated protocol to optimize cardiac motion compensation. An iodinated contrast agent (Iohexol, 350 mg I/mL) was administered at a volume of 70 mL, followed by a 40 mL saline chaser at a flow rate of 5.5 mL/s via antecubital venous access. SmartPrep bolus tracking was employed to ensure optimal contrast timing. The z-axis coverage was 157.5 mm, encompassing the entire heart from above the carina to below the diaphragmatic surface. A non-contrast scan was performed prior to contrast-enhanced CCTA to allow for coronary calcium scoring. Images were reconstructed across 11 temporal phases, each corresponding to a range of R-R interval values from -5% to 106%. Slice thickness was 0.625 mm with matching intervals, and the in-plane resolution was 0.23 x 0.23 mm. The mean effective radiation dose, calculated according to European guidelines for multi-slice CT, was 14.93 mSv (DLP: 1066.66 mGy·cm, using a conversion factor of 0.014 mSv/mGy·cm). Coronary plaque quantification and Agatston scoring were performed using GE's cardiac analysis software (AW VolumeShare 7), specifically the SmartScore module. As illustrated in Fig 2, our processing pipeline for cardiac CT data follows a sequential workflow beginning with retrospective ECG-gated image acquisition. The methodology proceeds through four key analytical stages: First, anatomical structure segmentation is performed, including multi-phase reconstruction and skeletal structure delineation. Second, cardiac and vascular reconstruction is conducted with detailed cardiovascular modeling of chambers, vessels, and pathological features. Third, EAT assessment quantifies the distribution and characteristics of pericardial fat deposits. Finally, dynamic modeling and motion analysis are achieved through temporal registration and displacement vector field analysis. This integrated approach enables a comprehensive evaluation of cardiac anatomy, function, and associated tissue characteristics, supporting both advanced clinical applications and research objectives.

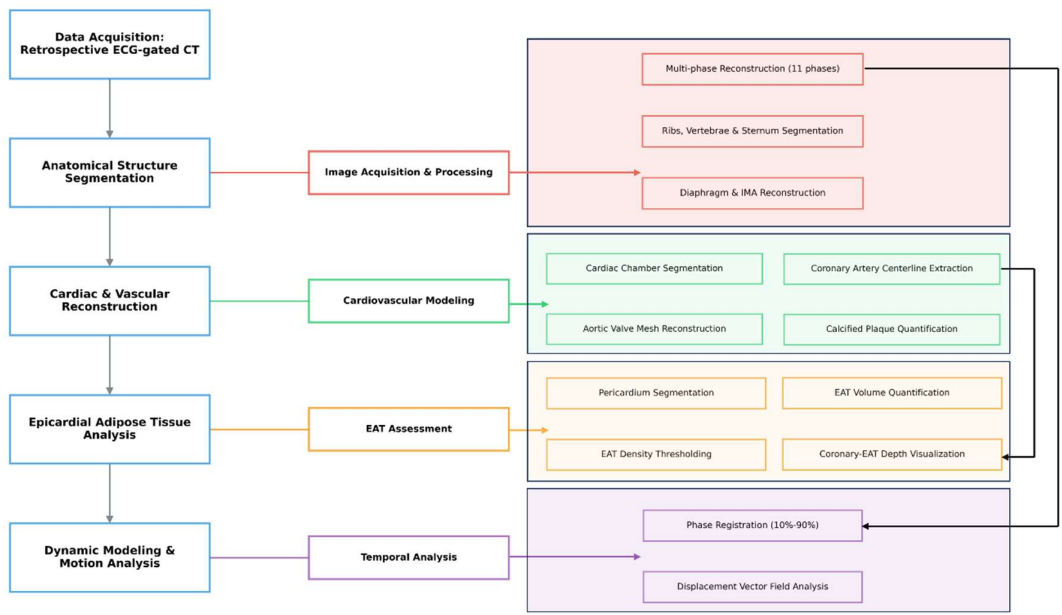

**Figure 2** Comprehensive cardiac CT imaging pipeline for cardiovascular analysis.

First, we implemented our approach using MONAI[14], an open-source framework for deep learning in healthcare imaging, utilizing its Auto3DSeg project (a component of the MONAI framework) for automated 3D segmentation. We segmented and reconstructed the ribs, vertebrae, sternum, diaphragm, and internal mammary artery from end-diastolic phase (70% R-R interval) CT images. Next, we applied HeartDeformNet[15] to automatically segment the left ventricle, right ventricle, left atrium, right atrium, left ventricular myocardium, aorta, and pulmonary artery across all phases of the cardiac cycle. A slightly modified version of DeepCarve[16] was used for aortic valve mesh reconstruction. The primary modification involved using a template from an end-diastolic segmented left ventricle model, which contained fewer distorted elements. An experienced cardiologist manually marked and recorded arterial contours on planes perpendicular to the vessel centerline using SimVascular-v2023.03.27[17], thus achieving segmentation of the coronary arteries from the 70% phase image. Subsequently, we imported nine cardiac phases (10–90%) and registered them to the selected 70% phase. This registration provided the transformations needed to adjust the 3D surface model points of the coronary arteries. These transformations were then used to generate a time-dependent sequence of 3D coronary artery models (Fig 3), completing the process of dynamic modeling.

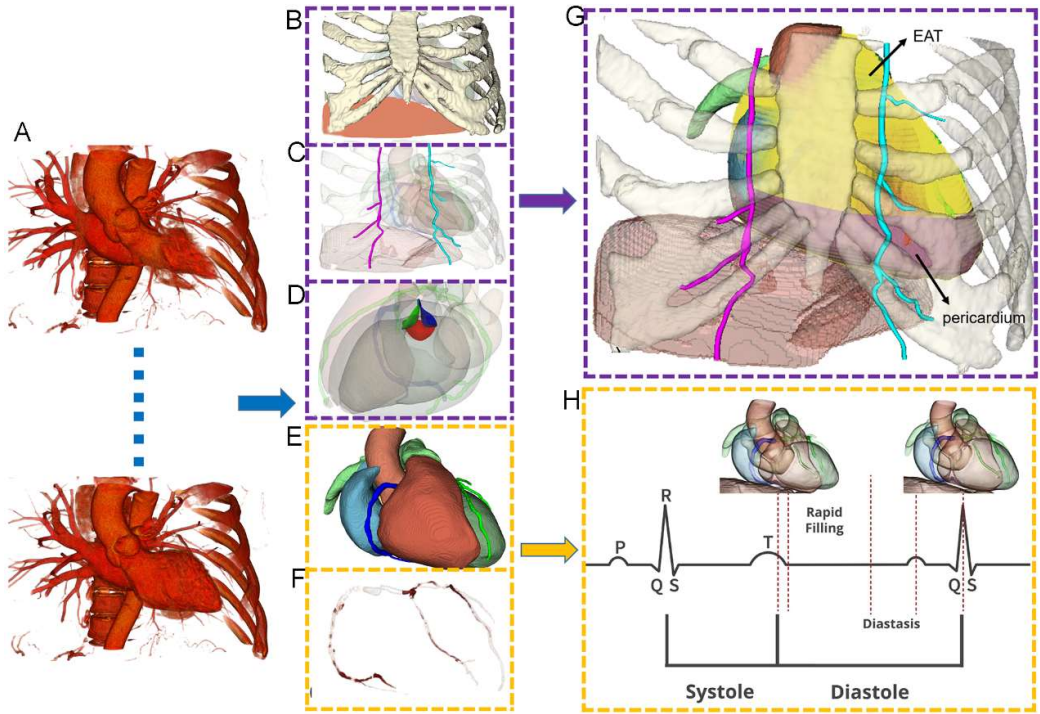

**Figure 3** Technical Visualization of Cardiac CT Analysis Results. (A) ECG-gated sequential cardiac images showing the heart at different phases of the cardiac cycle. (B) 3D reconstruction of the thoracic cage and diaphragm. (C) Internal mammary artery visualization and anatomical relationships. (D) Detailed aortic valve structural analysis. (E) Complete heart mesh model displaying chambers and great vessels. (F) 3D reconstruction of coronary arteries. (G) Complete scene view with labeled EAT and pericardium. (H) Dynamic model sequence of the cardiac cycle correlated with ECG waveform.

Second, for the visualization of calcified plaques, we followed the Agatston method[18]. We first identified contiguous calcium voxels within the 3D space of the binarized calcium image. Three quantitative indices—Agatston score, mass score, and volume score—were calculated for the major branches of the coronary arteries. In the calcified regions, medium and heavily calcified plaques were highlighted in red. Fiz et al.[19] classified plaque calcium density into tertiles: light (130–210 HU), medium (211–510 HU), and heavy (>510 HU) (Fig 4).

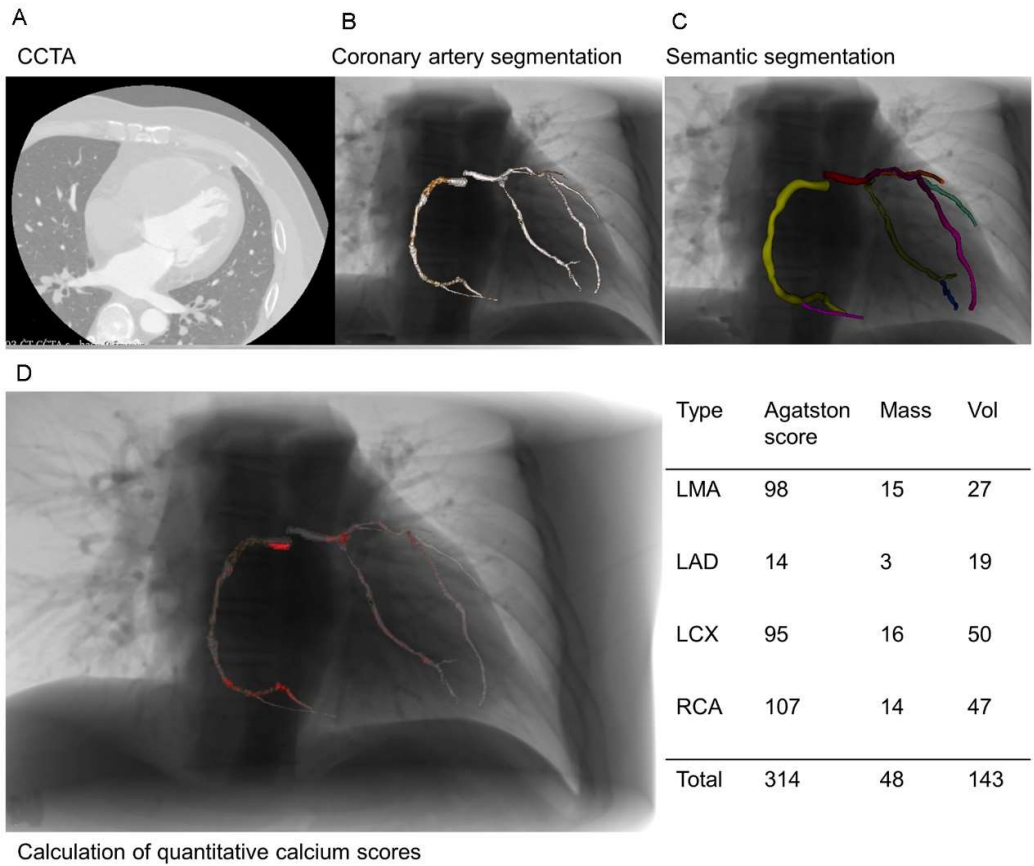

| Type | Agatston score | Mass | Vol |
|------|------|------|------|
| LMA | 98 | 15 | 27 |
| LAD | 14 | 3 | 19 |
| LCX | 95 | 16 | 50 |
| RCA | 107 | 14 | 47 |
| Total | 314 | 48 | 143 |

**Figure 4** Calculation and visualization of quantitative indices for coronary artery calcification. (A) Original CCTA image. (B) Coronary artery segmentation showing the contours of major coronary vessels. (C) Semantic segmentation with different coronary artery branches marked in distinct colors. (D) Quantitative analysis of calcified plaques, with visualization of calcification areas (marked in red) on the left, and a table on the right showing quantitative results including Agatston score, calcium mass, and volume for each coronary artery branch.

Third, to visualize the depth of coronary artery travel, we modeled EAT to assess how deeply the coronary arteries are embedded within it. EAT measurements were performed using a semi-automated procedure. Expert analysts segmented the pericardium using 3D Slicer v5.2.2[20] in a sequential, slice-by-slice manner. EAT was measured from the bifurcation of the pulmonary trunk to the apex of the heart. Manual pericardial contours were drawn on every 5–10 slices, followed by interpolation and corrections where necessary. Contiguous voxels with attenuation values between -190 and -30 Hounsfield units were used to define and quantify EAT[21]. A fusion display combining EAT and coronary artery models was used to present depth information, visualized through variations in coronary artery fluoroscopy (Fig 5).

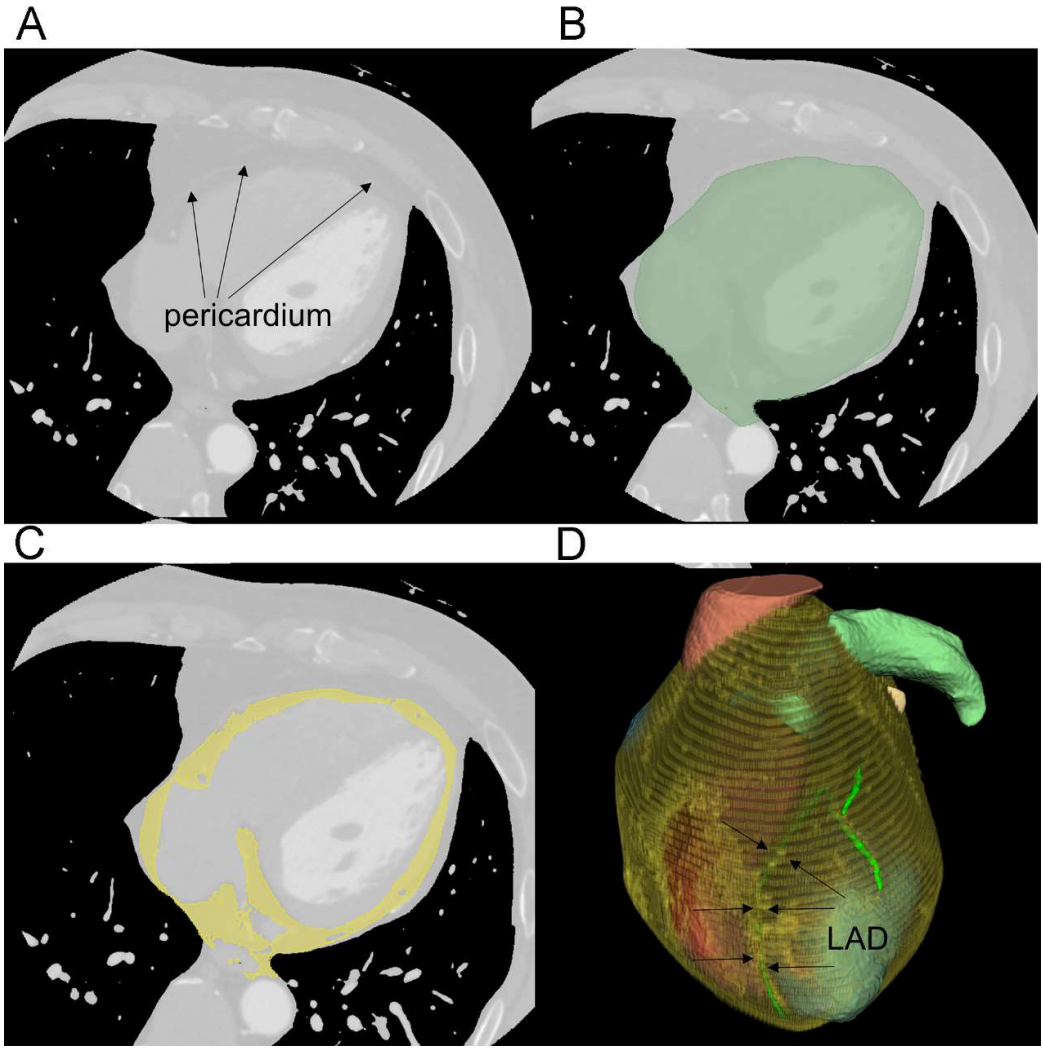

**Figure 5** Proposed EAT segmentation method. (A) The pericardium appears as a thin white line in the coronary CTA image. (B) The region encompassing the pericardium is first segmented by manual labeling (green region). (C) The threshold method is used to segment the adipose tissue within the pericardial mask (yellow region). (D) The volume rendering result of the segmented EAT.

Finally, when adhesions exist between the pericardium and the heart, the motion of the two structures tends to be synchronized. In contrast, the absence of adhesions results in noticeable differences in their relative motion. To observe this, we dynamically reconstructed both the pericardium and the outer cardiac tissue. Nine cardiac phases (10–90%) were imported and registered to the selected 50% phase.

This registration utilized the symmetric diffeomorphic algorithm proposed by Avants

et al.[22], generating a sequence of transformations represented as a time-varying 3D

displacement field. This sequence was then applied to deform the pericardium and

EAT (Fig 6).

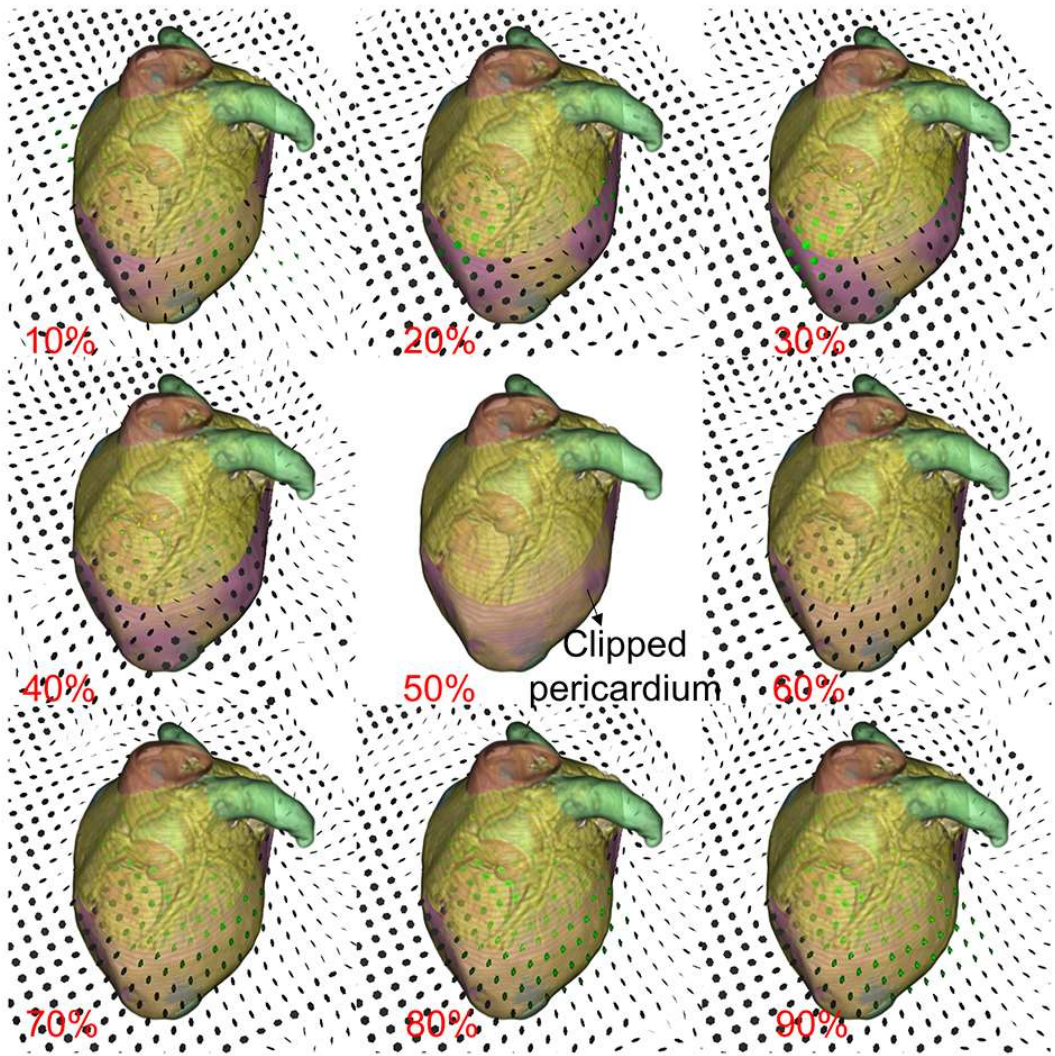

**Figure 6** Relative motion of pericardium and EAT. The pericardium is cut open to help observe the inner EAT structures. The corresponding deformation field of the cutting plane is visualized using arrows. Each arrow represents the displacement vector, indicating the direction and magnitude of displacement originating from its starting point on the plane.

## Evaluation of outcomes

### Evaluation of naked-eye holographic display

The outcome parameters were assessed qualitatively. Thirteen cardiac surgery clinicians evaluated various aspects of the holographic display, including display quality, realism of 3D spatial positioning, identification of calcified plaques, perception of coronary artery alignment depth, realism of dynamic visualization, and detection of pericardial adhesions. Evaluations were conducted using a Likert scale, based on preoperative and intraoperative data from 14 patients scheduled for CABG. In addition, the researchers assessed participants' attitudes toward the system's usefulness in preoperative planning and clinical education, as well as their willingness to adopt the system in practice. A score of 5 represented the most positive evaluation, while a score of 1 indicated the least favorable response (Multimedia Appendix 1: Questionnaire 1).

### Evaluation of pericardial adhesions

Pericardial adhesions were classified as either 0 or 1, where 0 indicated no adhesions and 1 indicated significant adhesions requiring the removal of a substantial portion of the pericardium. The intraoperative score was derived from the patient's operative report. Adhesion severity in the holograms was visually assessed by an experienced cardiovascular surgeon trained in interpreting holographic representations of pericardial adhesions, including the significance of the color schemes. The surgeon was blinded to all patient information. To assess intra-rater reliability, the same evaluator performed a second blinded evaluation of the same hologram samples two months later.

## Statistical analysis

The Spearman rank correlation test was conducted using SPSS version 25 for Windows (SPSS Inc., Chicago, IL, USA) to evaluate the correlation between hologram-based adhesion scores and intraoperative adhesion scores. Statistical analysis of the holographic display evaluation scores was performed using Microsoft Excel (version 2021 MSO). All statistical analyses were descriptive. Continuous variables were summarized using the mean, standard deviation, minimum, median, and maximum values, while categorical variables were reported as counts and percentages.

# Results

All 13 cardiac surgery professionals who tested the hologram system were male. The group consisted of six attending physicians and seven senior doctors, with ages ranging from 34 to 54 years (median: 37 years).

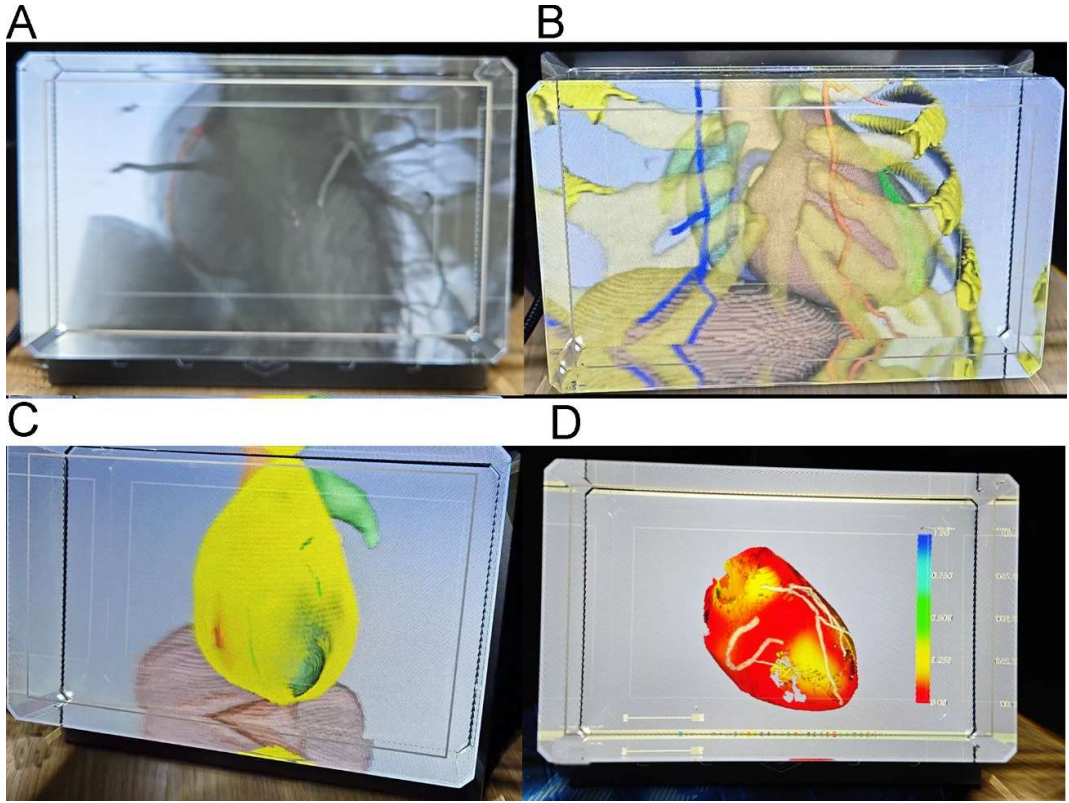

**Figure 7** Typical scenarios on holographic viewing system. (A) Dynamic 3D coronary artery scene segmented from CT data, medium and heavily calcified plaques are marked in red. (B) A 4D chest scene including the heart, coronary artery, internal mammary artery, ribs, vertebrae, sternum, and diaphragm. (C) Scenes demonstrating coronary depth. (D) Dynamic pericardium and cardiac motion scene.

When viewing the holograms from various angles (Fig 7, Multimedia Appendix 2: video 1), all participants—regardless of their level of expertise—agreed that the system offered a realistic visualization and allowed for accurate identification of the major coronary artery branches, the locations of severe stenosis, and calcified plaques. Additionally, by zooming and rotating the model, users could clearly view the spatial relationships among the heart, coronary arteries, and internal mammary artery within the thoracic cavity (median rating: 5/5, Table 1). However, two experts assigned scores of 2 and 3, respectively, for the identification of coronary stenosis and calcified plaques. They expressed concern that CCTA data may not reliably discriminate coronary stenosis and noted that intravascular contrast enhancement could impact plaque identification, potentially leading to inaccurate results. To assess inter-rater reliability, Fleiss' kappa coefficients were calculated for each evaluation item. The results indicated substantial to almost perfect agreement among evaluators across all assessment criteria ($\kappa$ ranging from 0.68 to 0.85). The highest agreement was observed for comparisons with traditional 2D displays ($\kappa = 0.85$) and dynamic display fidelity ($\kappa = 0.82$), reflecting strong consensus regarding the system's superiority over conventional visualization methods. While all criteria demonstrated substantial agreement, coronary calcified plaque visualization received the relatively lowest kappa score ($\kappa = 0.68$), suggesting some variability in expert perception of this feature. Overall, the high Kappa values confirm the reliability of the evaluation results and strengthen the validity of the positive assessments across all measured dimensions.

**Table 1**: Holographic experience reviews and attitude towards (future) use of holographic imaging system

| | Mean rating (SD) | Median rating (1–5) | Inter-rater agreement ($\kappa$) |
|---|---|---|---|
| Spatial perception | 4.78 (0.4) | 5 | 0.79 |
| Coronary calcified plaque | 4.28 (1.0) | 5 | 0.68 |
| Coronary depth | 4.71 (0.5) | 5 | 0.76 |
| Dynamic display fidelity | 4.78 (0.4) | 5 | 0.82 |
| Pericardial adhesions | 4.64 (0.6) | 5 | 0.73 |
| Basic tools (rotate, zoom) | 4.71 (0.5) | 5 | 0.79 |
| Compared to traditional 2D screens | 4.85 (0.4) | 5 | 0.85 |
| Overall evaluation | 4.71 (0.5) | 5 | 0.74 |
| Helpful for preoperative planning | 4.57 (0.5) | 5 | 0.71 |
| Beneficial for clinical teaching | 4.78 (0.4) | 5 | 0.81 |
| Useful for patient education | 4.64 (0.5) | 5 | 0.72 |

Anterior descending superficial myocardial bridges were observed in 2 of the 14 CABG patients based on CTA images. Evaluators perceived the depth trajectory of the coronary arteries within the EAT by noting a progressive transition in the holograms, from a blurred proximal segment to a clearer mid-to-distal segment. This trend corresponded with intraoperative findings, where anastomotic selection could not avoid the bridged region due to limitations in internal mammary artery length (Fig 8). Depth perception scores for the coronary arteries were consistently rated as "agree" or "strongly agree," with a median score of 5/5.

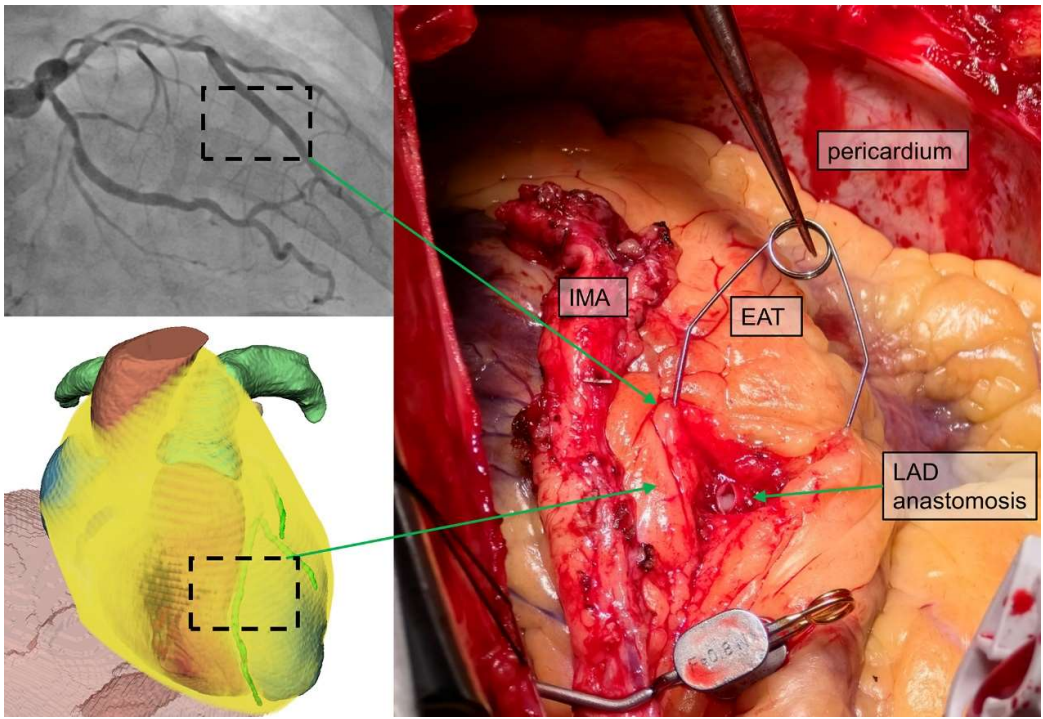

**Figure 8** The intraoperative situation of patient No. 9 shows that the anterior descending branch anastomosis proposed before surgery is located deep in the EAT, and the location of the anterior descending branch alignment is straight in the preoperative coronary angiography, implying that there might be myocardial bridges. The hologram produced results consistent with the intraoperative situation and the diagnostic results of the preoperative imaging.

Conventionally, images are viewed on a 2D screen following 3D reconstruction of the heart and coronary arteries using 3D Slicer software. To compare approaches, we asked evaluators to respond to the statement: "7. In obtaining the information needed for preoperative evaluation, it provides more intuitive and comprehensive insights compared to observing static three-dimensional cardiovascular CT images on a two-dimensional screen." The mean score was 4.85 (SD: 0.4), with 5 representing the most positive evaluation and 1 the least. Cardiac surgeons who participated in the evaluation agreed that the visualization method used in this study provided superior insights compared to traditional techniques. Key advantages included enhanced depth perception and more detailed cardiac modeling, particularly in representing coronary artery travel depth and pericardial adhesions.

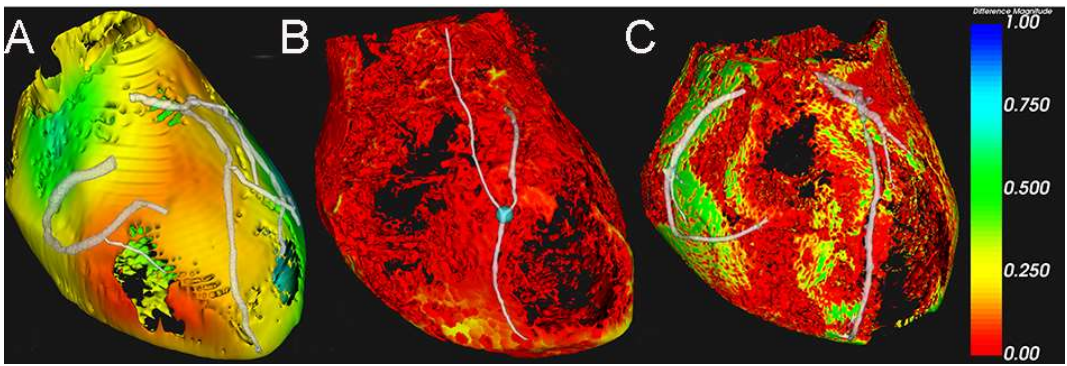

**Figure 9** Visualization images of pericardial adhesions. (A) Patient No. 6, scheduled for first-time CABG, exhibits significant relative motion between the pericardium and EAT, suggesting the absence of pericardial adhesions. (B) Patient No. 7, who underwent left ventricular outflow tract unblocking, mitral valve replacement, and CABG one year ago, demonstrates poor relative motion between the pericardium and EAT, indicating the presence of pericardial adhesions, consistent with intraoperative findings. (C) Patient No. 2, who had an aortic valve replacement one year prior, shows extensive pericardial adhesions in the operative report, but the hologram

pericardial adhesion score is rated as 0, suggesting no adhesions (Multimedia Appendix 3: video 2).

The hologram-based adhesion score showed a strong correlation with the intraoperative adhesion score (r = 0.786, P < 0.001, 95% confidence interval: 0.53–0.92; Table 2), confirming the validity of the data. All participants acknowledged the effective visualization of pericardial adhesions (median score: 5/5; Fig 9), and noted that the holographic display provided a more intuitive and 3D representation than traditional ultrasound images. A second blinded evaluation was conducted two months later to assess repeatability. The intra-rater reliability, measured by the Kappa coefficient, was 0.85, indicating strong consistency in the evaluation results.

**Table 2**: Holograms adhesion scores are determined by a surgeon blinded to the intraoperative observations. Intraoperative adhesion scores are determined by operative report.

|  | Intraoperative score | |
|  | 0 | 1 |
| --- | --- | --- |
| Hologram score | | |
| 0 | 14 | 1 |
| 1 | 0 | 6 |

# Discussion

## Principal findings

Holograms have demonstrated the ability to provide realistic depth perception, enhancing the intuitive understanding of a patient's 3D anatomy. However, the complexity of current hologram production methods has limited the widespread adoption of this technology in routine clinical practice. Additionally, existing head-mounted display devices are often uncomfortable and can hinder effective communication among healthcare team members. This study represents the first attempt to use real patient volumetric data—specifically 4D-CCTA—to generate high-quality, naked-eye-viewable, 3D, color dynamic holograms relevant to CABG surgery. This visualization tool eliminates the need for clinicians to rely on personal spatial imagination to synthesize relevant anatomical information. Instead, it offers a more intuitive, comprehensive, and vivid representation of the data required for preoperative CABG planning. The holograms can be simultaneously viewed and discussed by multiple surgeons or with patients, without the need for headgear. Unlike most existing cardiac and coronary models, this study introduces a specialized modeling framework tailored for CABG preoperative planning. It includes a dynamic representation of the pericardium, heart, and coronary arteries, all derived from the patient's 4D-CCTA data. Presenting volumetric patient data as dynamic, multicolor, and interactive holograms marks a significant advancement in cardiac medical imaging. This study outlines the fabrication process, structural framework, and preliminary feasibility assessment.

The preprocessing time for each patient dataset requires approximately 15 min, making it clinically feasible for use in preoperative planning. However, coronary artery segmentation remains the most time-consuming and technically challenging component—especially in patients with advanced CAD. In cases of significant vessel occlusion or calcification, which are common among CABG candidates, automated segmentation methods often require manual refinement to ensure clinical accuracy. This limitation is well recognized in the field and represents a key area for future algorithmic development.

In this study, the EAT, pericardium, and coronary arteries were segmented to characterize their relative motion and structural features, in order to extract clinically relevant information regarding coronary artery depth and pericardial adhesions. Pericardial adhesions increase both the difficulty and risk of minimally invasive or median sternotomy procedures and must be identified preoperatively. Currently, pericardial adhesions are typically evaluated using ultrasonography—which detects motion discrepancies between the pericardium and outer myocardium—or CT imaging, which assesses changes in pericardial density[23]. However, thickening of the pericardium on CT does not reliably correlate with adhesions, and although ultrasonography can reveal relative motion, it lacks 3D detail and fails to provide a comprehensive view of adhesion location relative to the coronary arteries. Another significant challenge is assessing the depth of coronary artery travel within the EAT. This is clinically important because intraoperative identification of deeply embedded coronary arteries can be difficult. Accurately obtaining this anatomical information preoperatively can significantly inform the selection of an anastomotic site during surgery. CCTA is currently the preferred modality for evaluating EAT[24]. Traditionally, EAT has been studied in terms of its volume or composition to explore its relationship with the progression of CAD[25]; however, these approaches offer limited insight into spatial relationships relevant for surgical planning. In this study, we dynamically fused models of epicardial fat and coronary arteries to represent their positional relationship and visualize the depth trends of the coronary arteries. Combined with a holographic display, this approach achieved the intended visualization effect. The accuracy of coronary artery depth representation was supported by comparison with intraoperative findings. Furthermore, the strong correlation between preoperative holographic assessments of pericardial adhesions and intraoperative scores provides preliminary validation of the system's clinical reliability.

## Comparison to prior work

The personalized dynamic heart model developed in this study provides more comprehensive anatomical information and a more realistic display, offering preliminary technical support for future hemodynamic analysis and intraoperative planning. In addition, for patients and their families who may lack medical expertise, holographic visualizations allow surgeons to clearly and accurately convey the necessity and potential risks of surgery. This approach may help build trust in the surgeon's decision-making and reduce anxiety stemming from the knowledge gap between doctors and patients.

Participants in the systematic evaluation of the holograms gave high ratings for both the quality of the display and the clinical relevance of the content. The features clinicians found most valuable included: 1) the ability to visualize spatial relationships between the coronary arteries and surrounding structures—particularly the depth of coronary arteries within EAT—which supports the planning of anastomotic sites; 2) the integration of multiple imaging modalities into a single, cohesive visualization platform, eliminating the need to mentally synthesize data from separate CT and angiographic images; and 3) the intuitive, glasses-free viewing experience, which facilitates collaborative discussion among the surgical team by enabling multiple clinicians to simultaneously view and interact with the same anatomical structures from different angles. These features directly address several clinical challenges in CABG planning that are not adequately resolved by conventional imaging methods, contributing to the positive reception of the holographic approach among the surgical team.

**Limitations**

First, our initial assessments were subjective and qualitative in nature and, therefore, not amenable to rigorous statistical analysis. Although clinician feedback was generally positive, we did not use validated quantitative metrics to compare the utility of the holographic visualization with conventional imaging techniques. Future studies should incorporate standardized assessment tools and objective performance measures to quantitatively evaluate the impact of holographic visualization on surgical planning and clinical outcomes.

Second, we did not perform quantitative analyses of calcified plaques, coronary artery travel depth, or pericardial–cardiac motion differentials. While we strived to maintain data fidelity throughout the visualization pipeline, it is important to recognize that the segmentation process inherently introduces variability and potential distortion. This may be influenced by factors such as the choice of segmentation algorithms, parameter settings, and manual corrections. Although experienced radiologists reviewed all segmentations to ensure clinical accuracy, this oversight does not entirely eliminate the risk of error. These limitations may have affected the accuracy of the spatial relationships depicted in our holograms.

Third, we did not quantify segmentation error or validate the geometric accuracy of our models against the original source data, which represents a significant limitation of the current methodology. Future work should include rigorous validation of segmentation accuracy and quantification of model fidelity relative to the source CT data to ensure clinical reliability.

Fourth, due to the limitations of current CT imaging, coronary stenosis still requires coronary angiography as the clinical "gold standard." While our approach integrates information from both CCTA and angiography, it does not yet match the diagnostic accuracy of invasive angiography for stenosis assessment. Future iterations could address this limitation by incorporating computational fluid dynamics to better evaluate the functional severity of stenosis.

Fifth, this study represents only a preliminary feasibility assessment of holographic visualization. We did not conduct quantitative or comparative studies of procedures performed using conventional imaging alone versus those supported by holographic assistance. Such comparative studies are essential to determine the practical utility, effectiveness, and broader applicability of holographic imaging in real-world clinical settings.

## Conclusions

This study outlines the structural framework of a visualization tool specifically designed for preoperative planning of CABG procedures. By leveraging actual patient-derived volumetric data, the system generates high-quality, clinically relevant, dynamic holograms that deliver true depth perception, detailed coronary stenosis and calcification information, alignment depth data, and 4D spatial insights—offering an entirely new visualization experience. Clinician feedback was overwhelmingly positive, confirming the tool's feasibility for preoperative planning. Additionally, this system shows promise for use in medical education, including surgical demonstrations for interns and preoperative education for patients.

## Future perspectives

The potential impact of holographic technology on medical imaging is substantial, particularly in the context of minimally invasive surgery. These procedures involve small incisions and a limited field of view, making a clear understanding of patient-specific spatial anatomy crucial. In future studies, we plan to further investigate the use of holographic imaging systems for preoperative planning in minimally invasive cardiac surgery. This includes integrating interactive tools and hemodynamic analyses and exploring the possibility of overlaying holograms onto the real intraoperative environment with interactive manipulation for real-time surgical guidance.